\documentclass[useAMS,usenatbib]{mn2e}
\usepackage{amsmath}
\usepackage{amssymb}
\usepackage{txfonts}
\usepackage{color}
\usepackage[breaklinks, colorlinks, citecolor=blue, linkcolor=blue, urlcolor=blue]{hyperref}
\usepackage{parskip}
\usepackage{graphicx}
\usepackage[usenames,dvipsnames]{xcolor}
\usepackage{multirow} 
\usepackage{soul} 

\definecolor{darkgreen}{rgb}{0.0, 0.5, 0.0}

\newcommand{\LCDM}{\Lambda\text{CDM}}
\newcommand{\M}{\ensuremath{{\rm M}}}
\newcommand{\Bc}{\ensuremath{\rm B^{(c)}}}
\newcommand{\Bd}{\ensuremath{\rm B^{(d)}}}
\newcommand{\Sc}{\ensuremath{\rm S^{(c)}}}
\newcommand{\Sd}{\ensuremath{\rm S^{(d)}}}
\newcommand{\Vc}{\ensuremath{\rm V^{(c)}}}
\newcommand{\Vd}{\ensuremath{\rm V^{(d)}}}

\newcommand{\Tregc}{\ensuremath{\rm T_{reg}^{(c)}}}
\newcommand{\Tregd}{\ensuremath{\rm T_{reg}^{(d)}}}
\newcommand{\Tirrc}{\ensuremath{\rm T_{irr}^{(c)}}}
\newcommand{\Tirrd}{\ensuremath{\rm T_{irr}^{(d)}}}

\newcommand{\s}{\ensuremath{\rm S}}
\newcommand{\V}{\ensuremath{\rm V}}
\newcommand{\Treg}{\ensuremath{\rm T_{reg}}}
\newcommand{\Tirr}{\ensuremath{\rm T_{irr}}}
\newcommand{\T}{\ensuremath{\rm T}}

\newcommand{\OmK}{\Omega_K}

\newcommand{\OmM}{\Omega_{\mathrm{m}}}
\newcommand{\OmL}{\Omega_{\Lambda}}
\newcommand{\sigmaSH}{\ensuremath{(\sigma_S/H)_0}}
\newcommand{\sigmaVH}{\ensuremath{(\sigma_V/H)_0}}
\newcommand{\sigmaTH}{\ensuremath{(\sigma_T/H)_0}}
\newcommand{\sigmaTregH}{\ensuremath{(\sigma_{T,\rm reg}/H)_0}}
\newcommand{\sigmaTirrH}{\ensuremath{(\sigma_{T,\rm irr}/H)_0}}

\title[Testing isotropy with the CMB]{A framework for testing isotropy with \newline the cosmic microwave background}
\author[Saadeh et al.]{Daniela Saadeh$^{1}$\thanks{daniela.saadeh.13@ucl.ac.uk}, Stephen M. Feeney$^{2}$, Andrew Pontzen$^{1}$, Hiranya V. Peiris$^{1}$ 
\newauthor and Jason D. McEwen$^{3}$\\
$^{1}$Department of Physics and Astronomy, University College London, London WC1E 6BT, U.K.\\
$^{2}$Astrophysics Group, Imperial College London, Blackett Laboratory, Prince Consort Road, London SW7 2AZ, U.K.\\
$^{3}$Mullard Space Science Laboratory (MSSL), University College London, Surrey RH5 6NT, U.K.}

\begin{document}

\maketitle

\begin{abstract}
{We present a new framework for testing the isotropy of the Universe using cosmic microwave background data, building on the nested-sampling \texttt{ANICOSMO} code. Uniquely, we are able to constrain the scalar, vector and tensor degrees of freedom alike; previous studies only considered the vector mode (linked to vorticity). We employ Bianchi type VII$_h$ cosmologies to model the anisotropic Universe, from which other types may be obtained by taking suitable limits. In a separate development, we improve the statistical analysis by including the effect of Bianchi power in the high-$\ell$, as well as the low-$\ell$, likelihood. To understand the effect of all these changes, we apply our new techniques to WMAP data. We find no evidence for anisotropy, constraining shear in the vector mode to $\sigmaVH < 1.7 \times 10^{-10}$ (95\% CL). For the first time, we place limits on the tensor mode; unlike other modes, the tensor shear can grow from a near-isotropic early Universe. The limit on this type of shear is $\sigmaTregH < 2.4 \times 10^{-7}$ (95\% CL).}\\
\end{abstract}

\section{Introduction}

The standard cosmological model assumes that space is homogeneous and isotropic on large scales. Observational data, particularly measurements of the cosmic microwave background (CMB), allow this assumption to be tested quantitatively.
When homogeneity and isotropy are assumed from the outset, the cosmological solutions to Einstein's field equations are described by Friedmann-Lema\^{i}tre-Robertson-Walker (FLRW) metrics. By relaxing the requirement for isotropy, one is instead led to Bianchi metrics \citep{Bianchi_originale,EllisMaccallum69}. The basis for testing isotropy is therefore to consider which of these backgrounds better describes our Universe. In the limit that departure from isotropy is small, the observed fluctuations in the CMB are well approximated by the sum of a deterministic Bianchi template and the stochastic contribution from the inflationary $\LCDM$ cosmological model.

Testing isotropy has received considerable attention since \emph{Wilkinson Microwave Anisotropy Probe} \citep[WMAP;][]{WMAP_ILC} full-sky maps became available. \citet{Jaffe_et_al_2005} found a correlation between WMAP temperature data and a pattern induced by the Bianchi VII$_h$ model; employing the new background also improved the fit to the temperature power spectrum \citep{Jaffe_et_al_2005, McEwen2006, Bridges2007}, suggesting that the Universe indeed departs from isotropy. However, the same authors pointed out that the best-fit Bianchi template is characterized by cosmological parameters (for example a large negative curvature) that are inconsistent with other available observations. 

\citet{Jason} subsequently introduced \texttt{ANICOSMO}, a tool for robust statistical analysis of the effects of an anisotropic background on the CMB. This code has been employed for a number of studies of the type of Bianchi models considered by \citet{Jaffe_et_al_2005}, most recently using \textit{Planck} data \citep{Planck_background_2013,Planck_background_2015}. In these analyses, in which the parameters for the background and stochastic components were required to be mutually consistent, no preference was found for anisotropy. However, the tests only took into account two out of a total five degrees of freedom of the Bianchi anisotropy, and thus did not allow an upper limit to be placed on anisotropy in general.

\citet{P&C11} presented a systematic linearization, from which the most general anisotropies that respect homogeneity can be expressed as a set of non-interacting modes on an isotropic background. To date, no statistical analysis is available for the additional degrees of freedom highlighted by this analysis, so that a true test of universal isotropy is lacking. Some of the previously-unconstrained modes are expected to be the most compatible with probes such as nucleosynthesis and CMB polarization, both of which are particularly sensitive to early-time anisotropy \citep{Pontzen09}. 

In this work, we create a framework for testing all modes simultaneously and validate it in the VII$_h$ case. This requires a number of changes to the previous approaches. We use the WMAP 9-year \citep{WMAP_9_yr} dataset previously considered by \citet{Jason} so that we can isolate the impact of these changes on the posterior distributions and the Bayesian evidence ratios with respect to $\LCDM$. The {\it Planck} analyses \citep{Planck_background_2013,Planck_background_2015} did not include a high-$\ell$ likelihood, and hence do not provide a full setting for comparison. In a future paper, we will extend the framework further to apply the new methods to {\it Planck} data, including polarization which is expected to be highly constraining \citep{P&C07}.

The paper is structured as follows: we introduce our method in Sec.~\ref{section:anisotropy} and \ref{section:statistical_analysis}, where we respectively discuss how we model anisotropy and the statistical analysis we perform to look for its signatures. In Sec.~\ref{section:Results} we present results from our analysis of WMAP 9-year data, and conclude in Sec.~\ref{section:Conclusions}.

\begin{figure}
 \centering
 \includegraphics[width=\columnwidth]{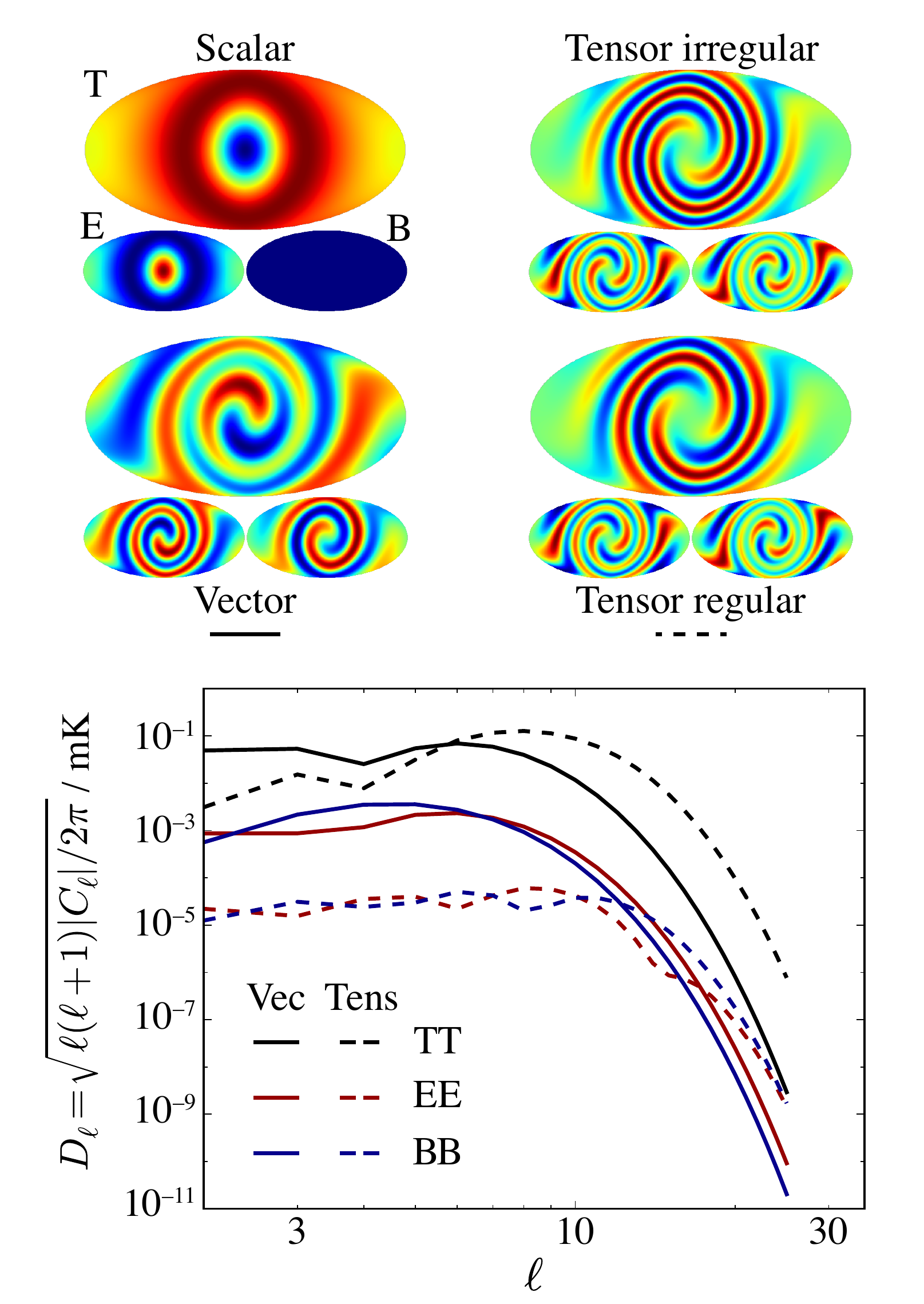}
 \caption{\textit{Maps:} Example scalar, vector, regular and irregular tensor patterns induced in the CMB temperature (upper panels) and polarization (lower panels, $E$- and $B$-mode to the left and right) for $( \OmM,\OmL,x) = (0.27, 0.7, 0.62)$. These maps were produced with the \texttt{ABSolve} code developed for this analysis. \textit{Plot:} Vector and regular tensor power spectra for $( \OmM,\OmL,x) = (0.27, 0.7, 0.62)$ and $(\sigma_V/H)_0=1\times10^{-9}$ (solid line), $(\sigma_{T,\rm reg}/H)_0=5\times10^{-6}$ (dashed line): significantly smaller values of the vector amplitude today lead to comparable temperature signals, and larger polarization signals, compared to that of the tensors. } \label{Fig:SVT_Dls}
\end{figure}

\section{Implementing anisotropy} \label{section:anisotropy}
In this Section, we describe the framework underlying \texttt{ABSolve} (\emph{Anisotropic Boltzmann Solver}), a Boltzmann-hierarchy code we have developed to compute the deterministic CMB temperature and polarization perturbations induced by a Bianchi background. We first discuss how anisotropy is modelled within the code in Sec.~\ref{section:shear_decomposition}, before presenting the computational details of the calculation in Sec.~\ref{section:computational_stuff}.

\subsection{The anisotropy degrees of freedom} \label{section:shear_decomposition}

\texttt{ABSolve} can compute the anisotropy-induced perturbations for Bianchi types\footnote{Bianchi `types' are subclasses into which Bianchi cosmologies are divided, grouping the different, inequivalent, ways for a 3-space to be homogeneous. Accessible reviews on Bianchi models may be found in \citet{Ellis_RC_book} and \citet*{Dynamical_systems_book}; also see \cite{P&C11} for a different approach.} I, V, VII$_0$ and VII$_h$. These cosmologies are sufficient to take into account all the open and flat Bianchi models that are close to isotropy, and therefore compatible with observations \citep{P&C11}. The closed Bianchi type IX only induces power at $\ell=2$, making it difficult to constrain: it is therefore currently not included.\footnote{Bianchi I also only induces power in the quadrupole, but it arises as the natural flat limit of Bianchi V.} In this paper, we focus on the Bianchi type VII$_h$, together with its flat limit VII$_0$.

Anisotropy is characterized quantitatively by means of the shear tensor $\sigma_{\mu\nu}$, which describes the deformation experienced by fluid elements in the Universe during anisotropic expansion; the shear scalar $\sigma$ is defined by $\sigma^2 \equiv \sigma_{\mu\nu} \sigma^{\mu\nu} / 2$.  To model the evolution of shear, we follow \citet{P&C11} in decomposing the anisotropic expansion into a set of five non-interacting modes on an underlying isotropic (FLRW) background. Each mode corresponds to a degree of freedom in the shear tensor. The deterministic perturbations induced in the CMB transform like scalars (1 d.o.f.), vectors (2 d.o.f.) or tensors (2 d.o.f.) under rotations around a preferred axis of the Bianchi symmetry, and are labelled accordingly throughout this work. In particular we write the magnitude of the shear associated with each component as $\sigma_S$, $\sigma_V$ and $\sigma_T$ respectively by considering their evolution independently; when all modes are combined one may show that the total shear obeys $\sigma^2 = \sigma_S^2 + \sigma_V^2 + \sigma_T^2$. Currently, published constraints on departures from isotropy use the approach of \cite{Barrow85} and consequently consider only $\sigma_V$, i.e., the two Bianchi VII$_h$ vector modes \citep[e.g.,][]{Jaffe_et_al_2005,Jason,Planck_background_2013,Planck_background_2015}. 

{The evolution of anisotropy is dictated by the Einstein equations. We assume the contents of the Universe can be described as a sum of perfect fluids corresponding to radiation, matter and dark energy. Under these assumptions, scalar and vector modes exhibit a fast decay ($\sigma_V$, $\sigma_S \propto a^{-3}$, where $a$ is the scale-factor), linking small levels of anisotropy today to larger levels at recombination. Extrapolating backwards to the Big Bang, eventually the shear becomes comparable to the Hubble parameter, at which point the linear decomposition of \citet{P&C11} ceases to apply. We refer to such behaviour in this paper as ``irregular'', since it cannot be immediately reconciled with a near-isotropic early-Universe implied by the inflationary scenario. The steep decay also gives rise to a high degree of polarization in the CMB \citep{P&C07}. }

{Solutions for the tensor modes can behave in this irregular way; but there is also a growing solution \citep{CollinsHawking73} allowing observable anisotropy to emerge from a near-isotropic early Universe. We term this behaviour ``regular'', since it can more easily be fitted into the modern cosmological paradigm, although fine-tuning of inflation is still required for shear to reach an observable amplitude at the present day. Only the tensor degrees of freedom, which have not been tested previously, can exhibit this behaviour. Scalar and vector modes also possess a second solution, but it can be removed by a coordinate transformation and therefore has no physical effect. }

{The maps in Fig.~\ref{Fig:SVT_Dls} show, from upper left to lower right, CMB anisotropies imprinted by scalar, irregular tensor, vector and regular tensor modes.\footnote{For animations of the Bianchi pattern for varying cosmological parameters, see DOI \href{http://dx.doi.org/10.5281/zenodo.48654}{\tt 10.5281/zenodo.48654}.} The inset panels show the polarization. For the case of the vector and regular tensor modes, the Bianchi power spectra are plotted underneath as solid and dashed curves respectively. Black, red and blue lines show temperature, $E$-mode and $B$-mode polarization power spectra respectively. }

{The magnitude of the shear in these cases has been chosen to produce a similar rms temperature anisotropy amplitude of approximately $75\,\mu\mathrm{K}$. For the case of the vector modes a present-day shear (normalized to the isotropic Hubble expansion rate $H_0$ to form a dimensionless quantity) of $\sigmaVH \simeq 10^{-9}$ is sufficient. However for the regular tensor modes, this amplitude must be considerably higher, $\sigmaTregH \simeq 5 \times 10^{-6}$, because the steep scaling with redshift is absent. We can therefore immediately anticipate that constraints on present-day anisotropy in regular tensor modes will be considerably weaker than the corresponding constraints for the vector modes.}

\begin{table}
\caption{Summary of the parameters defining the Bianchi morphology, amplitude and orientation. The amplitudes are expressed in terms of the shear scalar normalized to the Hubble parameter. In addition to the parameters tabulated below, there is the discrete choice of handedness.} \label{Tab:Bianchi_params}
 \begin{tabular}{|l|l|}
 \hline
 \multicolumn{2}{|c|}{Morphology}\tabularnewline
  \hline
  $\OmM$ & matter density \tabularnewline
  $\OmL$ & dark energy density \tabularnewline
  $x$ & rotation scale of shear principal axes \tabularnewline
  \hline
 \multicolumn{2}{|c|}{Signal amplitude} \tabularnewline
 \hline
  $\sigmaSH$ & \parbox[t]{6cm}{amplitude of scalar modes} \tabularnewline
  $\sigmaVH$ & \parbox[t]{6cm}{amplitude of vector modes} \tabularnewline
  $\sigmaTregH$ & \parbox[t]{6cm}{amplitude of regular tensor modes} \tabularnewline
  $\sigmaTirrH$ & \parbox[t]{6cm}{amplitude of irregular tensor modes} \tabularnewline
  \hline
 \multicolumn{2}{|c|}{Orientation} \tabularnewline
 \hline
  $\alpha$, $\beta$, $\gamma$ & \parbox[t]{6cm}{Euler angles defining the orientation of the Bianchi pattern} \tabularnewline
  \hline
 \end{tabular}
\end{table}

Having decomposed the shear into $\sigmaSH$, $\sigmaVH$ and $\sigmaTH$ there are a number of further degrees of freedom to be considered. First, the specific initial conditions for the tensor mode are placed into either the decaying, irregular solution (in which case we will refer to $\sigma_{T,\mathrm{irr}}$) or the growing, regular solution ($\sigma_{T,\mathrm{reg}}$). Additionally, the Bianchi morphology is set by the matter and dark energy densities today $\{\OmM, \OmL\}$ and the parameter $x$, which regulates the scale over which the shear principal axes rotate in the vector and tensor modes (see \citealt{CollinsHawking73,Barrow85} for a formal definition). The orientation relative to the Galaxy is determined by three Euler angles, labelled $\alpha$, $\beta$ and $\gamma$. There is then a further freedom to rotate the tensor shear contribution relative to that of the vectors by an angle $\gamma_{VT}$; however, we will not consider vector and tensor modes simultaneously in the present work, and this freedom therefore does not enter. 
Finally, Bianchi vector and tensor modes have a handedness, so for these models parity must also be specified. For reference, all the Bianchi parameters employed in this work are summarized in Table \ref{Tab:Bianchi_params}.

\subsection{Computational aspects} \label{section:computational_stuff}

The deterministic contribution to the CMB is obtained by computing the Boltzmann equation for photons in the context of a given Bianchi background: the complete set of equations can be found in \citet{P&C07}, except for the shear evolution, derived by \citet{P&C11}.  Our code \texttt{ABSolve} is written in Python and Cython. Run times vary considerably across the parameter space, but typically take a few seconds on one 2.6 GHz core.

The code starts by computing the recombination history using {\texttt{RECFAST 1.3}}\footnote{\url{http://www.astro.ubc.ca/people/scott/recfast.html}.} \citep{RECFAST,RECFAST2}. The Boltzmann integration starts at redshift $z_{\text{start}}=1500$ with zero initial Bianchi power; the power quickly builds from the shear-induced temperature quadrupole and scatters into the $E$-mode polarization quadrupole. Anisotropies are subsequently advected to smaller scales and partially converted into $B$-mode polarization due to free-streaming effects after recombination. We have checked for one example model\footnote{$\{ \Omega_b h^2 = 0.022, \Omega_c h^2 = 0.11, x=0.5, \OmM=0.27, \OmL=0.7,  \sigmaSH=1\times10^{-9}, \sigmaVH=1\times10^{-9}, \sigmaTregH=1\times10^{-6}, \sigmaTirrH=1\times10^{-7}, \gamma_{VT} =0 \}$} that the Bianchi pattern is unchanged if integration is started significantly earlier, at $z_{\text{start}}=1800$ or $z_{\text{start}}=2000$. This reflects the structure of the Boltzmann equation, where anisotropy is built by the shear tensor but damped by Thomson scattering through a viscous-friction term proportional to anisotropy. A limit equilibrium exists in this setting that is reached for any initial redshift sufficiently above that of recombination. A possible future improvement to {\tt ABSolve} would be to include a more refined treatment of $z_{\text{start}}$ that fixes its value based on the model parameters (in particular, taking into account the sensitivity of recombination to the baryon and dark matter physical densities). 

The maximum multipole used to characterize the Bianchi pattern, $\ell_{\mathrm{max}}$, must be chosen carefully to avoid missing small-scale power. We implemented a test that compares the power around a trial $\ell_{\mathrm{max}}$ with the power at the quadrupole; if this test fails, integration is repeated up to a much larger multipole. {Further details of this procedure and its importance are given in Sec.~\ref{section:small_scales}; typical values are $\ell_{\mathrm{max}}=200$ as an initial guess and $\ell_{\mathrm{max}}=1000$ subsequently. }

{We must also avoid edge effects that can propagate errors from the hierarchy truncation. We found empirically that we needed to
extend the calculated hierarchy to $\ell_{\mathrm{trunc}}=\ell_{\mathrm{max}}+50$ to obtain converged results at $\ell_{\mathrm{max}}$. To prevent instabilities from developing, we additionally apply a Fermi-Dirac damping of the form
\begin{equation}
 D(\ell)=\frac{1}{\exp\left[{\ell-(\ell_{\mathrm{trunc}}-10)}\right]+1}
\end{equation}
to the hierarchy\footnote{Technically, we implement this by multiplying the advection coefficients in equations (48), (52), (53) of \cite{P&C07} by $D(\ell)$. The chosen form for $D(\ell)$ is heuristically motivated, rather than following from any analytic approximation: we verified that it leads to convergence in all the examined cases.}, which damps the power at $\ell_{\mathrm{trunc}}-10=\ell_{\mathrm{max}}+40$ to prevent advected power reflecting at the unphysical boundary.}

Once our code has produced output in harmonic space, we use \texttt{healpy}\footnote{\url{http://healpix.sourceforge.net}} \citep{Healpix} to perform the required Euler rotations and to produce maps when needed.

\vspace{-0.3cm}
\section{Statistical Framework}  \label{section:statistical_analysis}
In this Section, we introduce the statistical framework required for constraining all possible modes of background anisotropy using CMB data. We start with an overview in Sec.~\ref{section:analysis_overview} before discussing the choice of priors in Sec.~\ref{section:priors}. Sec.~\ref{section:tests} presents an illustrative analysis as a guide to the interpretation of our results. Finally, in Sec.~\ref{section:small_scales} we show how the new analysis improves constraints on the background anisotropy by employing information from smaller scales (higher $\ell$) than considered in previous analyses.

\vspace{-0.2cm}
\subsection{Analysis overview} \label{section:analysis_overview}

In the models under study, stochastic $\LCDM$ fluctuations are superimposed on a Bianchi background. In this setting, the observed CMB data, ${\mathbf d}$, are assumed to consist of a stochastic, Gaussian $\LCDM$ component, ${\mathbf s}$, a deterministic Bianchi component, ${\mathbf b}$, and Gaussian instrumental noise, ${\mathbf n}$:
\begin{equation}
{\mathbf d} = {\mathbf s}(\Theta_{\Lambda\text{CDM}}) + {\mathbf b}(\Theta_{\rm B}) + {\mathbf n} .
\end{equation}
Here $\Theta_{\Lambda\text{CDM}}=\{\Omega_bh^2,\Omega_ch^2,\OmL,\Omega_K,n_{\mathrm{s}},A_{\mathrm{s}},\tau\}$ is a vector of parameters from the standard $\LCDM$ framework: baryon and dark matter physical densities $\Omega_bh^2$ and $\Omega_ch^2$ (where $H_0=100h$ km/s/Mpc), dark energy and curvature densities $\OmL$ and $\Omega_K$, scalar spectral index and power amplitude $n_{\mathrm{s}}$ and $A_{\mathrm{s}}$, and optical depth to reionization $\tau$. $\Theta_{\rm B}$ is a vector of the Bianchi parameters summarized in Table \ref{Tab:Bianchi_params} (see Sec.~\ref{section:shear_decomposition}). The partial overlap between $\Theta_{\rm B}$ and $\Theta_{\Lambda\text{CDM}}$ is discussed in more detail in Sec.~\ref{section:priors}.

The likelihood function $P(\mathbf{d}|(\Theta_{\text{B}},\Theta_{\Lambda\text{CDM}}),\M) = \mathcal{L}\left(\Theta_{\text{B}}, \Theta_{\Lambda\text{CDM}}\right)$ takes the form of a Gaussian with mean {$\mathbf{b}$} set by the deterministic Bianchi template and covariance matrix $\mathbf{C}$ set by the stochastic $\LCDM$ component and instrumental noise properties
\begin{equation}
 \mathcal{L}(\Theta_{\text{B}}, \Theta_{\Lambda\text{CDM}})\propto \frac{1}{\sqrt{|\mathbf{C}(\Theta_{\Lambda\text{CDM}})|}}\exp{\left[-\chi^2(\Theta_{\Lambda\text{CDM}}, \Theta_{\text{B}})/2\right]},\label{Eq:lowl_likelihood}
\end{equation}
where
\begin{equation}
 \chi^2(\Theta_{\Lambda\text{CDM}},\Theta_{\text{B}})=\left[\mathbf{d}-\mathbf{b}(\Theta_{\text{B}})\right]^{\dagger}\mathbf{C}(\Theta_{\Lambda\text{CDM}})^{-1}\left[\mathbf{d}-\mathbf{b}(\Theta_{\text{B}})\right].
\end{equation}
The data and Bianchi template may be expressed in either pixel or harmonic space.

{Limits on anisotropy can be considered as either a parameter-estimation or a model-comparison problem. In the former case, the product of the prior and likelihood gives the posterior probability density for different shear amplitudes. In the latter case} one must compute and compare each model's Bayesian evidence. The Bayesian evidence is defined as the probability of obtaining the data given a model $\M$, integrating over all of $\M$'s parameters $\Theta$, i.e., the marginal likelihood
\begin{equation}
 E = P(\mathbf{d}\,|\,\M) = \int d\Theta\, P(\mathbf{d}\,|\,\Theta,\M) \, P(\Theta\,|\,\M) .
\end{equation}
Here $P(\Theta\,|\,\M)$ is the a priori probability for $\Theta$ and $P(\mathbf{d}\,|\,\Theta,\M)$ the likelihood function. The evidence can be used to estimate the relative probability of two models $\M_1$ and $\M_2$, given available data
\begin{equation}
 \frac{P(\M_1\,|\,\mathbf{d})}{P(\M_2\,|\,\mathbf{d})} = \frac{P(\mathbf{d}\,|\,\M_1)}{P(\mathbf{d}\,|\,\M_2)}\frac{P(\M_1)}{P(\M_2)}=\frac{E_1}{E_2}\frac{P(\M_1)}{P(\M_2)}.
\end{equation}
If, as assumed in this work, no a priori knowledge is available that favours one model over the other (i.e., $P(\M_1)=P(\M_2)$), then the probability ratio equals the evidence ratio: $P(\M_1\,|\,\mathbf{d}) / P(\M_2\,|\,\mathbf{d}) = E_1 / E_2$. The log-Bayes factor $\ln(E_1/E_2)$ will be used in Sec.~\ref{section:Results} to assess the degree by which one model is favoured over the other.

In order to evaluate the probability distributions described above, we have integrated {\tt ABSolve} into the \texttt{ANICOSMO} package \citep{Jason}. {\tt ANICOSMO} uses the nested sampler {\tt MultiNest}\footnote{\url{http://ccpforge.cse.rl.ac.uk/gf/project/multinest/}} \citep{MultiNest1, MultiNest2, MultiNest3} to explore the parameter space of each model and hence efficiently calculate its Bayesian evidence. At each sampled point, the theoretical mean of the CMB data is evaluated using {\tt ABSolve}, and the covariance (set by the stochastic $\LCDM$ fluctuations) is calculated using power spectra produced by {\tt CAMB}~\citep{CAMB}\footnote{\url{http://camb.info/}}.

{When applied to masked data, the likelihood~\eqref{Eq:lowl_likelihood} is difficult to evaluate at high $\ell$ since the mask becomes problematic in harmonic space while the covariance becomes problematic in pixel space. It is therefore necessary to adopt an approximate power-spectrum-based likelihood at high $\ell$ \citep[typically $\ell>32$; e.g.,][]{ILC_problems,Page07}. In previous versions of {\tt ANICOSMO}, including those used by the Planck Collaboration \citep{Planck_background_2013,Planck_background_2015}, only the low-$\ell$ part of the likelihood has been modified to take into account the Bianchi template. This results in a good estimate of the overall likelihood provided that the power in the Bianchi component is negligible compared to the stochastic power in the high-$\ell$ modes. However, Bianchi models have two physical scales (a spiralling and curvature scale controlled by $x$ and $\Omega_K$ respectively); when either of these is sufficiently small relative to the horizon, high-$\ell$ power can be significant; see Sec.~\ref{section:small_scales}.}

For this reason, {we complement the low-$\ell$ likelihood~\eqref{Eq:lowl_likelihood} with an approximate modification to the high-$\ell$ likelihood using the summed contributions of the Bianchi and $\LCDM$ power spectra.} For fluctuations around an anisotropic Bianchi background the power spectrum does not provide lossless data compression, but in the limit that the Bianchi signal is subdominant relative to the $\LCDM$ component, the approximation is valid. See Appendix \ref{section:Appendix_Cl_approximation} for further details.

In test runs on simulated maps (see Sec.~\ref{section:tests}), full-sky information is available at all multipoles. In these cases, we use the likelihood function in Eq.~\eqref{Eq:lowl_likelihood} up to $\ell=400$, without an additional high-$\ell$ component. The absence of a high-$\ell$ likelihood constraining the temperature power spectrum in the damping tail makes the recovered constraints on test $\LCDM$ parameters less stringent. This is acceptable, since we only need to verify our ability to recover the Bianchi template parameters.

\subsection{Models} \label{section:priors}

\begin{table}
\caption{Models tested in this work. The $^{\mathrm{(d)}}$ superscript, where present, refers to the phenomenological `decoupled' models; physical `coupled' models are indicated with $^{\mathrm{(c)}}$. $\s$, $\V$, $\Treg$, $\Tirr$ refer respectively to Bianchi scalar, vector, regular and irregular tensor modes. \label{Tab:shorthand} }
 \begin{tabular}{|l|l|}
 \hline
  Notation & Model\tabularnewline
  \hline
  $\LCDM$ & pure $\LCDM$, no Bianchi component\tabularnewline\tabularnewline
  $\Sc$ & \parbox[t]{6cm}{Bianchi VII$_h$/VII$_0$ scalar modes, `coupled' model} \tabularnewline\tabularnewline
  $\Sd$ & \parbox[t]{6cm}{Bianchi VII$_h$/VII$_0$ scalar modes, `decoupled' model} \tabularnewline\tabularnewline
  $\Vc$ & \parbox[t]{6cm}{Bianchi VII$_h$/VII$_0$ vector modes, `coupled' model}\tabularnewline\tabularnewline
  $\Vd$ & \parbox[t]{6cm}{Bianchi VII$_h$/VII$_0$ vector modes, `decoupled' model} \tabularnewline\tabularnewline
  $\Tregc$ & \parbox[t]{6cm}{Bianchi VII$_h$/VII$_0$ regular tensor modes, `coupled' model}\tabularnewline\tabularnewline
  $\Tregd$ & \parbox[t]{6cm}{Bianchi VII$_h$/VII$_0$ regular tensor modes, `decoupled' model}\tabularnewline\tabularnewline
  $\Tirrc$ & \parbox[t]{6cm}{Bianchi VII$_h$/VII$_0$ irregular tensor modes, `coupled' model}\tabularnewline\tabularnewline
  $\Tirrd$ & \parbox[t]{6cm}{Bianchi VII$_h$/VII$_0$ irregular tensor modes, `decoupled' model}\tabularnewline\tabularnewline
  \hline
 \end{tabular}
\end{table}

As described in Sec. \ref{section:analysis_overview}, parameters related to density appear in both $\Theta_{\mathrm{B}}$ and $\Theta_{\LCDM}$. A physically meaningful analysis must have self-consistent matter and dark energy densities $\OmM$ and $\OmL$ when calculating the background and stochastic perturbation contributions: such analyses are referred to as Bianchi `coupled' runs in the following, or $\Bc$. Our upper limits on anisotropy will be derived in these settings. However, to connect with the early analyses that found evidence in favour of Bianchi cosmologies, we also test phenomenological models in which the two components are allowed to vary independently: these models are labelled `decoupled', or $\Bd$. 

\begin{table}
\begin{center}
\caption{Priors for all the model parameters, along with the search to which they are applied. Where no indication is given, the stated prior is applied in all cases: $\Bc$, $\Bd$ and pure $\LCDM$. See also Sec.~\ref{section:priors}.} 
 \begin{tabular}{|c|c|c|c|}
\hline 
Parameter & Prior Range & Prior type & Models \tabularnewline
\hline 
$\Omega_bh^2$ & $[0.005,0.05]$ & uniform & \tabularnewline
$\Omega_ch^2$ & $[0.05,0.3]$ & uniform &  \tabularnewline
$\OmL$ & $[0,0.99]$ & uniform & \tabularnewline
$\Omega_K$ & $[10^{-5},0.5]$ & uniform & $\Bc$ \tabularnewline
$\OmK$ & 0 & fixed & $\Bd$ and $\LCDM$ \tabularnewline
$n_{\mathrm{s}}$ & $[0.9,1.05]$ & uniform & \tabularnewline
$A_{\mathrm{s}}$ & $[1,5]\times10^{-9}$ & log-uniform & \tabularnewline
$\tau$ & $[0.082,0.092]$ & uniform & \tabularnewline
\hline
$\OmM$ & $[0,0.99]$ & uniform & $\Bd$ \tabularnewline
$\OmL$ & $[0,0.99]$ & uniform & $\Bd$ \tabularnewline
$x$ & --- & --- & $\s$ \tabularnewline
$x$ & $[0.05,2]$ & uniform & $\V$, $\Treg$ and $\Tirr$ \tabularnewline
$\sigmaSH$ & $[-10^{-8},10^{-8}]$ & uniform & $\s$ \tabularnewline
$\sigmaVH$ & $[10^{-12},10^{-8}]$ & log-uniform & $\V$ \tabularnewline
$\sigmaTregH$ & $[10^{-12}$,$10^{-4}]$ & log-uniform & $\Treg$ \tabularnewline
$\sigmaTirrH$ & $[10^{-12}$,$10^{-4}]$ & log-uniform & $\Tirr$ \tabularnewline
$\alpha$ & $[0^{\circ},360^{\circ}]$ & uniform & $\Bc$ and $\Bd$ \tabularnewline
$\beta$ & $[0^{\circ},180^{\circ}]$ & sine-uniform & $\Bc$ and $\Bd$ \tabularnewline
$\gamma$ & --- & --- & $\s$ \tabularnewline
$\gamma$ & $[0^{\circ},360^{\circ}]$ & uniform & $\V$ \tabularnewline
$\gamma$ & $[0^{\circ},180^{\circ}]$ & uniform & $\Treg$ and $\Tirr$ \tabularnewline
\hline 
\end{tabular}\label{Tab:priors}
\end{center}
\end{table}

Table \ref{Tab:shorthand} lists the models considered in this work. {We use a Bianchi VII$_h$ model and allow the curvature to approach zero such that the VII$_0$ models are also naturally included.}
We separately test for the scalar ($\s$), vector ($\V$), regular ($\Treg$) and irregular ($\Tirr$) tensor degrees of freedom in turn; the priors that we adopt for the Bianchi and $\LCDM$ parameters are listed in Table \ref{Tab:priors}. 
For $\sigmaVH$, $\sigmaTregH$ and $\sigmaTirrH$, we adopt log-uniform priors so as to avoid setting a preferred scale for these parameters. This choice is not possible for $\sigmaSH$, which can take negative values in our parametrization: we adopt a uniform prior in this case.

\begin{figure}
 \centering
 \includegraphics[width=\columnwidth]{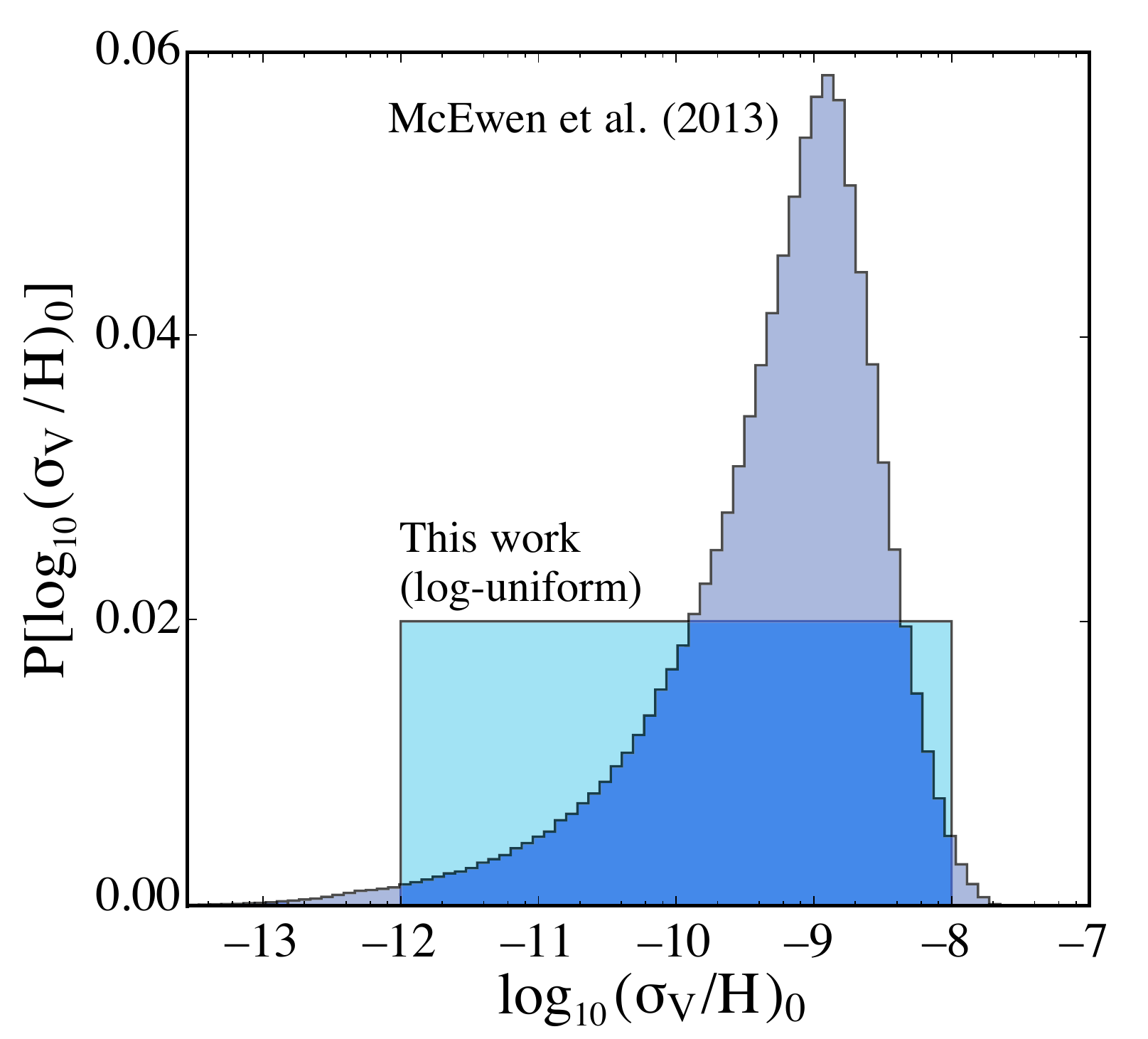}
\caption{Comparison of the prior on $\sigmaVH$ employed in this work (log-uniform) with that implied by \citet{Jason} in which a uniform prior is taken over $(\omega/H)_0$, $x$ and $\OmK$. The transformation between the spaces is described by Eq.~\eqref{Eq:wH_vs_sH}. The approximate range is comparable, but the prior of \citet{Jason} places considerable added emphasis on shear values around $\sigmaVH \simeq 10^{-9}$.
\label{Fig:prior_vorticity_vs_shear}}
\end{figure}

The expanded set of modes requires us to use a general parameterization that uses shear, rather than vorticity, to control the amplitude of anisotropy. Consequently our priors cannot be made identical to those in previous work. 
The link between expansion-normalized shear $\sigmaVH$ and vorticity $(\omega/H)_0$ is provided by the time-space component of the Einstein equations, which gives \citep{Barrow85}:
\begin{equation}
 \left(\frac{\omega}{H}\right)_0 = \frac{\sqrt{1+x^2\,\OmK}\,\sqrt{1+9\,x^2\,\OmK}}{6\,x^2\,(1-\OmK)}\left(\frac{\sigma_V}{H}\right)_0 \textrm{.} \label{Eq:wH_vs_sH}
\end{equation}
In \cite{Jason}, the prior on $(\omega/H)_0$ is uniform over the range $[0, 10^{-9}]$. However the appearance of parameters $x$ and $\Omega_K$ in the relationship~\eqref{Eq:wH_vs_sH} shows that the implied prior on $\sigmaVH$ is not uniform. By sampling points from the prior of \cite{Jason} and applying the mapping~\eqref{Eq:wH_vs_sH}, we established that the marginalized prior on $\log \sigmaVH$ in previous work is peaked around $\sigmaVH \simeq 10^{-9}$; see Fig.~\ref{Fig:prior_vorticity_vs_shear}. By contrast we assign equal a priori probability to all the scales in the range $[10^{-12},10^{-8}]$. As an example of the Bianchi signal strengths that are consequently included in the search, the lower and upper limits of our prior range correspond to rms temperature fluctuations of respectively $0.05$ and $500\,\mu\mathrm{K}$ at $x=0.3$.

The allowed ranges for $\sigmaTregH$ and $\sigmaTirrH$ need to be significantly wider than that on $\sigmaSH$ and $\sigmaVH$ because similar values of $\s$ or $\V$ and $\Treg$ or $\Tirr$ at recombination correspond to considerably larger values for the tensor shear amplitude today, especially in the case of $\Treg$: for the tensor cases, we allow values as large as $10^{-4}$.
Scalar modes possess rotational symmetry around a preferred axis, which makes them insensitive to the parameters $x$ and $\gamma$; they are consequently held fixed in the S runs. Similarly, because of the spin-2 tensor symmetry, the $\gamma$ angle is only required to vary in the range $[0^{\circ},180^{\circ}]$ in $\T$ runs, whereas it takes the full range $[0^{\circ},360^{\circ}]$ in $\V$ cases. The prior range for the optical depth to reionization, $\tau$, was chosen to match that of \cite{Jason} to ease comparison.

\subsection{Illustrative analysis} \label{section:tests}

\begin{figure}
 \centering
 \includegraphics[width=\columnwidth]{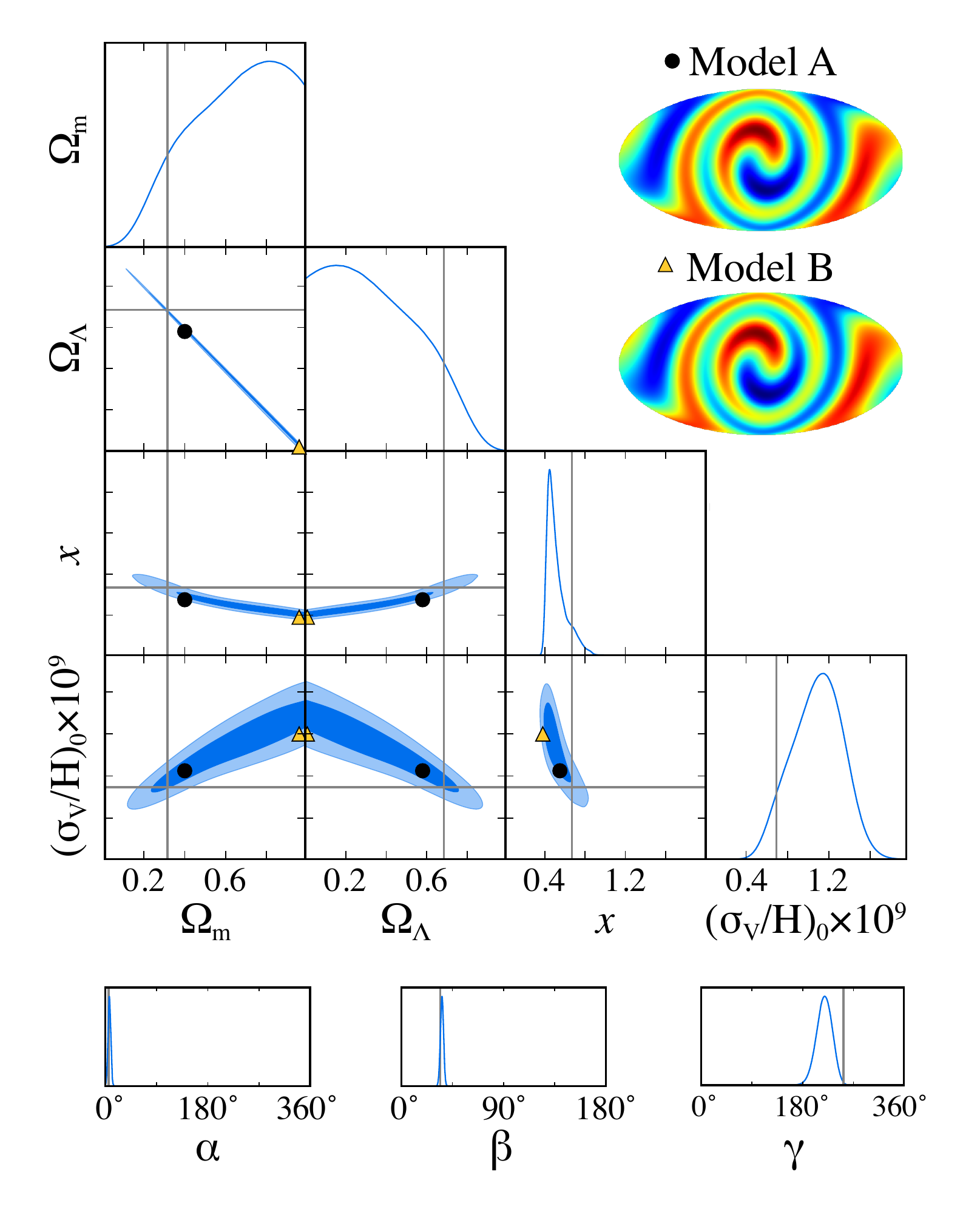}
 \caption{An illustration of the strong geometric degeneracy in the Bianchi parameter space. The triangle plot shows the recovered posterior distribution for a mock map containing a stochastic $\LCDM$ and underlying deterministic Bianchi pattern with $\OmM = 0.31$, $\OmL = 0.69$, $x=0.7$ and $\sigmaVH=0.7 \times 10^{-9}$. These input parameters are indicated by grey lines. The parameters are recovered but are strongly broadened by the geometric degeneracy arising from the projection of the Bianchi spiral onto the surface of last scattering. The two inset maps correspond to models near the ends of the recovered degeneracy as indicated by the dot (upper map) and triangle (lower map). The overall orientation of the pattern is constrained as seen in the sharply peaked marginalized posterior distributions of the Euler angles in the bottom row.} \label{Fig:Tests}
\end{figure}

We applied the analysis outlined in Sec.~\ref{section:analysis_overview} to several simulated maps of the CMB sky containing a Bianchi signal to check that the input parameters were recovered correctly. The mocks mimic stochastic $\LCDM$ fluctuations on top of a Bianchi background and were generated as follows: we computed the temperature power spectrum given a set of cosmological parameters through \texttt{CAMB} and obtained a realization of the corresponding Gaussian random field; we then added the resulting fluctuations linearly to a map containing a Bianchi template. The deterministic Bianchi contribution was calculated using the code from \cite{Pontzen09} as a blind test of the \texttt{ABSolve} implementation.\footnote{The \cite{Pontzen09} code is several times slower than \texttt{ABSolve} and its design decisions concerning timestepping and high-$\ell$ truncation are not suited to a statistical search.} We applied a Gaussian beam with FWHM of 1$^{\circ}$ to the maps and assumed instrumental noise to be negligible on the relevant scales. When applying our modified version of \texttt{ANICOSMO} to these mocks, we employed a likelihood function of the form in Eq.~\eqref{Eq:lowl_likelihood} up to $\ell=400$, with no supplementary high-$\ell$ likelihood.

Figure \ref{Fig:Tests} shows the recovered constraints in an example decoupled test run on a mock CMB map containing a Bianchi vector $\Vd$ signal. The recovered posterior is consistent with the input parameters, which are respectively $\OmM = 0.31$, $\OmL = 0.69$, $x=0.7$ and $\sigmaVH=0.7 \times 10^{-9}$. However, there is a strong degeneracy in the $\{\OmM,\,\OmL,\,x,\,\sigma/H\}$ dimensions; this reflects how the angular scale of the recovered pattern is approximately set by the Bianchi spiralling scale projected onto the last scattering surface. The inset temperature maps show how the input signal is mimicked extremely well along this degeneracy line. As a consequence, the marginalized posteriors on these parameters are broad.  The orientation of the pattern, defined by the Euler angles, is recovered well, with sharply peaked Gaussians around the input values, with the exception of $\gamma$ which is biased by $2.3\,\sigma$. We verified that the cause of this bias was a slight net rotation between the \cite{Pontzen09} maps which form the basis of the test and the \texttt{ABSolve} maps for the same input values. Given the more careful numerical choices described above, we believe the \texttt{ABSolve} results to be more robust.

Improved constraints on the Bianchi parameters can be obtained by employing complementary information to break the degeneracy: this is the case in $\Bc$ models, where $\LCDM$ fluctuations strongly limit the range of allowed matter and dark energy densities $\{\OmM,\,\OmL\}$, thereby also tightening the constraints on $x$ and $\sigma/H$. Taking into account the CMB polarization in addition to temperature would also partially break the model degeneracy. 

Our framework was additionally tested against mock {$\Sd$, $\Vc$, $\Vd$, $\Tregc$ and $\Tirrd$ maps}.

\subsection{The importance of small scales} \label{section:small_scales}

As discussed in Sec.~\ref{section:analysis_overview}, one novel feature of our improved framework is that it includes the effect of the Bianchi temperature fluctuations in the high-$\ell$ part of the likelihood. While some Bianchi models induce power that decays rapidly with $\ell$ and is negligible for $\ell>32$, there is a large part of our parameter space in which the high-$\ell$ corrections are significant. In particular, models with low values of $x$ or large negative curvatures have anisotropic features on strongly sub-horizon scales which project onto high $\ell$s.

\begin{figure}
 \centering
 \includegraphics[width=\columnwidth]{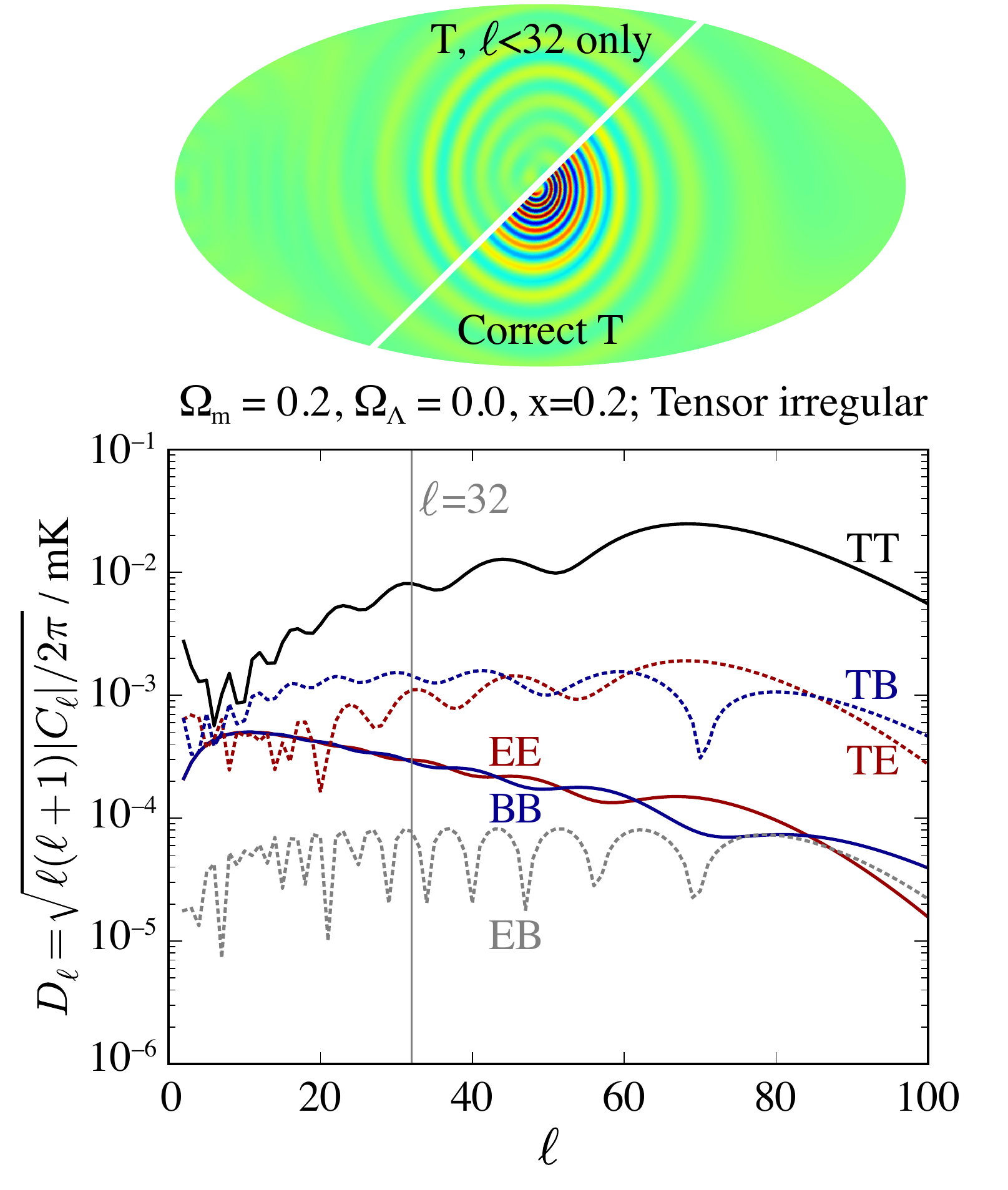}
 \caption{The effects of neglecting the small scales in the treatment of the Bianchi background. An \texttt{ABSolve} temperature map for the model $\{\OmM=0.2,\OmL=0,x=0.2\}$ was produced twice; first, using multipoles only to $\ell_{\mathrm{max}}=32$ (upper left portion of map) and second, using multipoles up to $\ell_{\mathrm{max}}=200$ (lower right portion of map). Most of the defining features in the Bianchi pattern are lost with the $\ell=32$ cut. The power spectra of the model are illustrated, with the $\ell=32$ cut highlighted with a vertical line, reinforcing how significant information is discarded if higher multipoles are not considered.} \label{Fig:small_scales}
\end{figure}

{Fig.~\ref{Fig:small_scales} illustrates a case where neglecting high-$\ell$ information is inappropriate. The split map shows how the intense alternating cold-and-hot spirals (lower-right portion) are lost when $\ell>32$ is ignored (upper-left portion). This is reflected in the Bianchi power spectrum $D_{\ell}$ which peaks at $\ell \simeq 70$. Without the high-$\ell$ information, models such as this will be inaccurately characterized.}

{Losing this information can equally cause false negatives and false positives. In the case of a true Bianchi universe characterized by a tight spiral, the statistical search will underestimate the likelihood around the correct parameters. Conversely, in the case of a pure $\LCDM$ universe, the statistical search will overestimate the likelihood of a tightly-spiralling feature. In particular, the final results of existing analyses must therefore spuriously favour low values of $x$ and overestimate the upper limit on $(\sigma/H)_0$. Our inclusion of the high-$\ell$ information will produce tighter and more robust limits on anisotropy. }

{For practical purposes we still need to apply a truncation to the Boltzmann hierarchy; since as $x \to 0$ the power is advected to arbitrarily high $\ell$, this generates a lower limit on the values of $x$ we can meaningfully consider.} We define $\ell^*$ to be the minimum multipole at which the average power over the range $\{\ell^*-10<\ell\leq\ell^*\}$ is significantly smaller than the power at the quadrupole according to the criterion\footnote{We consider the average $\frac{1}{10}\sum_{\ell=\ell^*-10}^{\ell^*} C_{\ell}$, as opposed to the power $C_{\ell^*}$ at a single multipole, to reduce the impact of localized dips in the power spectrum spuriously satisfying the constraint in Eq.~\eqref{Eq:lstar} at lower multipoles.}
\begin{equation}
 \frac{1}{10}\sum_{\ell=\ell^*-10}^{\ell^*} C_{\ell}<\frac{1}{100}C_{\ell=2}. \label{Eq:lstar}
\end{equation}
As discussed in Sec.~\ref{section:computational_stuff}, this test is performed after each integration. If $\ell^*$ exceeds $\ell_{\mathrm{max}} = \ell_{\mathrm{trunc}}-50$, the entire integration is repeated with a higher value of $\ell_{\mathrm{max}}$ up to 1000. Higher values of $\ell_{\mathrm{max}}$ become extremely slow to evaluate so we designed our priors to avoid models where $\ell^*$ exceeds $1000$; specifically, we exclude regions with $x<0.05$. This final threshold was chosen after calculating $\ell^*$ across a grid of 27,000 ($30 \times 30 \times 30$) models spanning the $\{ \OmM,\,\OmL,\,x \}$ unit cube at regular intervals. By this approach we also verified that $\ell^*$ rises towards small $x$ and large negative curvatures, as expected. We calculated that the implicit $\ell=32$ cut that has been applied previously mischaracterizes $16\%$ of the total cube. Therefore we expect significant changes to posteriors when our high-$\ell$ treatment is included.

\section{{Application to WMAP temperature data}}\label{section:Results}

{As a demonstration of the framework developed above,} we analyze the WMAP 9-year temperature data using a combination of the Internal Linear Combination (ILC) map for large scales \citep{WMAP_ILC} ($\ell \le 32$) with the $TT$ high-$\ell$ likelihood\footnote{\url{http://lambda.gsfc.nasa.gov/product/map/dr5/likelihood_get.cfm}} \citep{WMAP_9_yr} for small scales ($\ell>32$). The combination is chosen because it allows access to the full sky for the low-$\ell$ modes while avoiding the complex noise properties of the ILC on small scales \citep{ILC_problems}.
An earlier version of the ILC was the basis for finding a correlation with a Bianchi VII$_h$ template \citep{Jaffe_et_al_2005}.

Following the approach described in Sec.~\ref{section:analysis_overview}, the low-$\ell$ likelihood is specified by Eq.~\eqref{Eq:lowl_likelihood}. For the ILC, we employ a Gaussian beam with FWHM of 1$^{\circ}$ and assume that instrumental noise and residual foreground contamination is negligible at $\ell\leq32$. {The calculation is performed in harmonic space and no masking is applied.} The WMAP high-$\ell$ likelihood code models noise and beams internally. 

All the results discussed in this section were obtained by applying the priors listed in Table \ref{Tab:priors}. The complete posterior distributions and triangle plots are available from DOI \href{http://dx.doi.org/10.5281/zenodo.48653}{\tt 10.5281/zenodo.48653}.

This section is structured as follows: in Sec.~\ref{section:constraints}, we present the constraints we recover for the different anisotropy modes. In Sec.~\ref{section:evidence}, we compare the Bayesian evidence for Bianchi models and $\LCDM$. In Sec.~\ref{section:prior_effect}, we {discuss how our prior choices impact on the calculations compared to previous work.}

\subsection{Constraints on anisotropy} \label{section:constraints}

We tested the full anisotropy freedom of the Bianchi VII$_h$/VII$_0$ expansion using the WMAP 9-year data. As described above, our analysis considers the vector (vorticity) modes that have been studied previously, as well as new degrees of freedom that have not previously been included. Table \ref{Tab:results} summarizes the constraints we recover for the amplitudes of scalar, vector, regular and irregular tensor modes, as obtained when searching for $\Bc$ models.

\begin{figure}
 \begin{center}
  \includegraphics[width=0.5\textwidth]{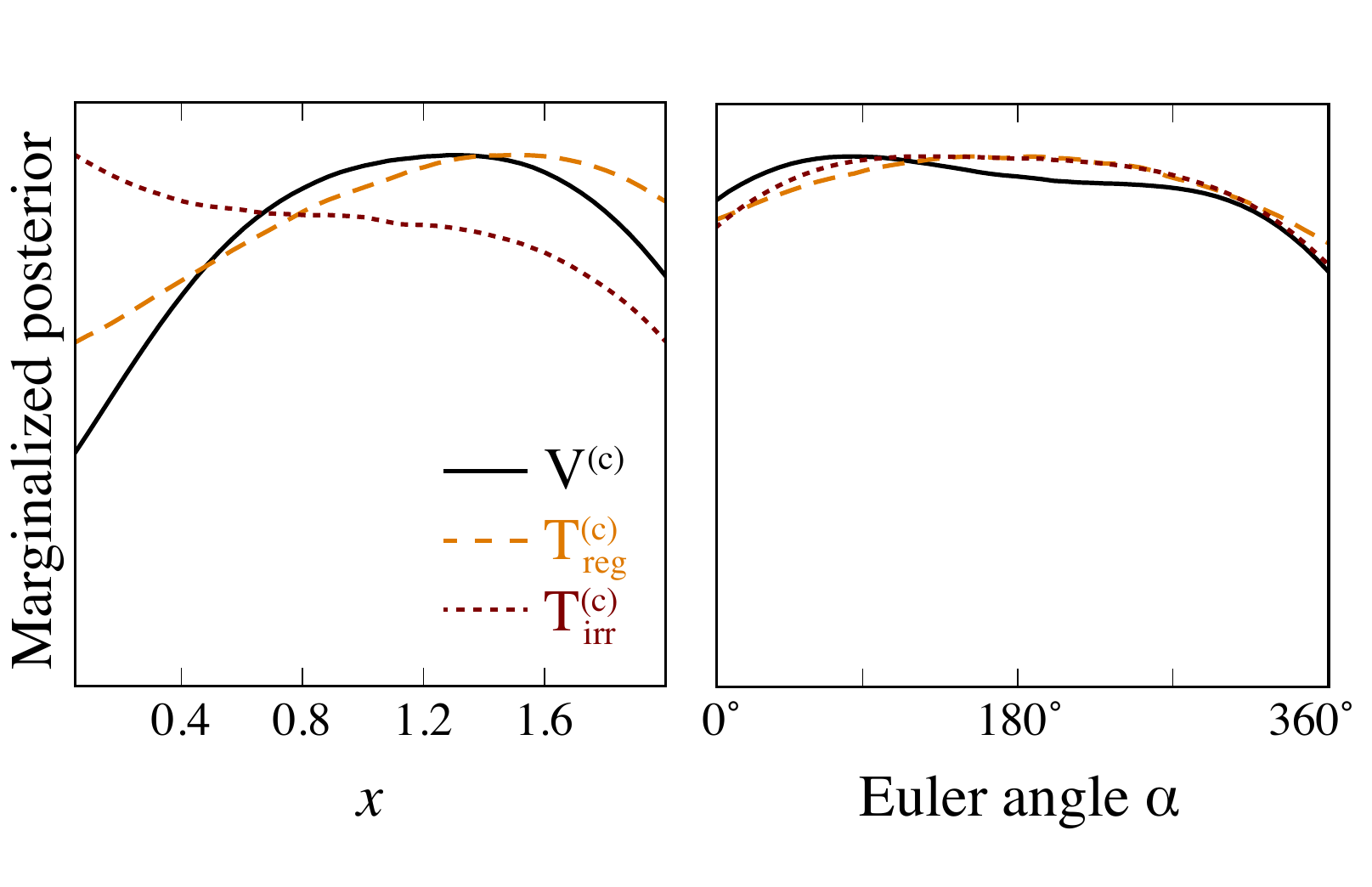}
  \caption{{Posterior distribution for the Bianchi $x$ parameter (left panel) and one of the Euler angles $\alpha$ (right panel) for $\Vc$, $\Tregc$ and $\Tirrc$ in left-parity models (right-parity models give similar results). In all cases the data do not strongly prefer any particular values, which reflects the lack of evidence for anisotropy in the data. A comparable analysis of the $\Vc$ case \citep{Jason} resulted in posteriors rising towards small $x$ (tight spirals); the difference can be traced to our improved analysis including high-$\ell$ information. } \label{Fig:x_posterior} }
 \end{center}
\end{figure}

For left-parity $\Vc$ modes, we obtain $\sigmaVH < 1.7\times10^{-10}$ (95\% CL).  $\OmM$ and $\OmL$ peak around concordance values, driven by the stochastic component. In Fig.~\ref{Fig:x_posterior} (left panel) we show how the Bianchi degree of freedom affecting the morphology, $x$, is largely unconstrained, with only the tightest spirals being marginally disfavoured. This suggests no overall preference by the data for specific Bianchi patterns; accordingly, the Euler angles are also unconstrained (see right panel of Fig.~\ref{Fig:x_posterior} for an example). Similar results hold for the right-parity $\Vc$ runs. 

In runs on $\Vd$, the Bianchi sub-case for which early evidence was found in the WMAP ILC map, we see a small preference in the parameters controlling the morphology (weakened by the degeneracy) and a sharper preference in the orientation, but, unlike in previous work, we only recover upper limits on the shear amplitude even in the presence of this considerable extra freedom. In Sec.~\ref{section:prior_effect}, we discuss how the difference in our result can be traced to the choice of priors.

For the case of V modes, it is possible to compute the universal vorticity $(\omega/H)_0$ through Eq.~\eqref{Eq:wH_vs_sH}. This allows constraints to be placed on $(\omega/H)_0$ from $x$, $\OmK$, and $\sigmaVH$; we obtain $(\omega/H)_0 < 1.6 \times 10^{-10}$ (95\% CL). However it must be stressed that this limit depends not only on the amplitude of the Bianchi signal, but also on the priors for $x$ and $\Omega_K$.

The three remaining degrees of freedom are constrained for the first time. Table \ref{Tab:results}  shows that the $\Tregc$ mode {(which we argued in Sec.~\ref{section:shear_decomposition} to be the best-motivated scenario from the standpoint of accommodating residual anisotropy within the standard cosmological paradigm) is constrained at a level three orders of magnitude weaker than $\Vc$ modes. This results from the regular behaviour which allows for relatively high levels of late-time anisotropy even when the early universe was near-isotropic (see Sec.~\ref{section:shear_decomposition}). For the other tensor solution $\Tirrc$, we obtain limits that are more closely comparable to the $\Vc$ case.} In all tensor cases, as in the vector cases, we find that $\OmM$ and $\OmL$ remain sharply peaked around concordance values and the Euler angles and Bianchi $x$ parameter display no strong preferences (Fig.~\ref{Fig:x_posterior}); in other words, the data do not support the existence of anisotropy in these modes.

\begin{table}
  \caption{Constraints on the Bianchi VII$_h$/VII$_0$ anisotropy (95\% CL).\label{Tab:results}}
 \begin{center}
   \begin{tabular}{|c|c|c|}
 \hline 
Mode & Parity & 95\% CL \tabularnewline
\hline
$\Sc$ & - & $-3.4 \times 10^{-10} < \sigmaSH < 3.8 \times 10^{-10}$ \tabularnewline
\rule{0pt}{3ex}
\multirow{2}{*}{$\Vc$} & Left & $\sigmaVH < 1.7 \times 10^{-10}$ \tabularnewline
                       & Right & $\sigmaVH < 1.6 \times 10^{-10}$ \tabularnewline
\rule{0pt}{3ex}
\multirow{2}{*}{$\Tregc$} & Left & $\sigmaTregH < 2.4 \times 10^{-7}$ \tabularnewline
                          & Right & $\sigmaTregH < 2.2 \times 10^{-7}$ \tabularnewline
\rule{0pt}{3ex}
\multirow{2}{*}{$\Tirrc$} & Left & $\sigmaTirrH < 2.4 \times 10^{-9}$  \tabularnewline
                          & Right & $\sigmaTirrH < 2.1 \times 10^{-9}$ \tabularnewline
\hline 
\end{tabular}
 \end{center}
\end{table}

The $\Sc$ modes are constrained to $-3.4 \times 10^{-10} < \sigmaSH < 3.8 \times 10^{-10}$ (95\% CL). Due to the scalar symmetry, the parameter $x$ has no effect and the orientation of the scalar pattern only requires two Euler angles to be defined, as it is rotationally invariant around a preferred axis; the data do not prefer any particular values for these two angles. Once again, concordance values are recovered for $\OmM$ and $\OmL$.
{The upper limit on $\Sc$ is slightly less stringent than that on $\Vc$ only because of the different prior shape on $\sigmaSH$ and $\sigmaVH$.} 

In summary, for all modes considered, the marginalized posterior distributions peak around the concordance values for the matter and dark energy densities, with little difference between the $\s$, $\V$, $\Treg$ and $\Tirr$ runs. {The data do not display any significant preference in the remaining parameters that control Bianchi morphology or orientation. In the specific case of the $x$ parameter, a distinct preference for low values has been found in previous work considering the same data \citep{Jason} but is absent in our analysis (Fig.~\ref{Fig:x_posterior}). This is a consequence of our refined treatment of small scales in the background modelling, which results in a more accurate assessment of the relative probability of Bianchi models with tightly wound spirals (Sec.~\ref{section:small_scales}). }

\subsection{Model comparison} \label{section:evidence}

Table \ref{Tab:Bayes_factor} shows the log evidence ratios of all examined models with respect to $\LCDM$.
The self-consistent $\Bc$ models are all strongly disfavoured compared to standard flat $\LCDM$. In \citet{Jason}, however, the Bianchi hypothesis had comparable evidence to $\LCDM$. This difference results from a combination of the improvements introduced in our method (particularly the treatment of the small scales, Sec.~\ref{section:small_scales}) and the choice of the prior on $\sigmaVH$ (Fig.~\ref{Fig:prior_vorticity_vs_shear}; Sec.~\ref{section:priors}). 
The updated prior choice also affects the evidence ratio for the $\Vd$ decoupled model, removing the preference found in \citet{Jason} for the left parity over the right one.

{$\s$ models stand out as they display a significantly smaller log-Bayes factor than the other degrees of freedom; however, this is a consequence of the uniform prior that we have to adopt for $\sigmaSH$ (see Sec.~\ref{section:priors}). The smallest shear amplitudes are favoured by the data but are given less weight by the uniform (rather than log-uniform) prior, so the evidence values are pushed down. To verify that this effect accounts for the apparent disfavouring of S models, we calculated the log evidence ratio for $\Vc$ and $\Vd$ with a uniform prior on $\sigmaVH$ $[0, 10^{-8}]$. The values become, respectively, $-6.0 \pm 0.2$ and $-1.7 \pm 0.2$ for the left parity, confirming that the uniform prior accounts for the down-weighting.}

\begin{table}
\caption{Log-Bayes factor for different Bianchi+$\LCDM$ models with respect
to standard flat $\LCDM$ (positive/negative values favour/disfavour
the addition of a Bianchi component).} \label{Tab:Bayes_factor}
\begin{center}
 \begin{tabular}{|c|c|c|c|}
 \hline 
Mode & Parity & $\Bc$ models & $\Bd$ models \tabularnewline
\hline
$\s$ & - & $-6.3 \pm 0.2$ & $-2.0 \pm 0.2$ \tabularnewline
\rule{0pt}{3ex}
\multirow{2}{*}{$\V$} & Left & $-3.4 \pm 0.2$ & $-0.1 \pm 0.2$ \tabularnewline
                       & Right & $-3.3 \pm 0.2$ & $0.0 \pm 0.2$ \tabularnewline
\rule{0pt}{3ex}
\multirow{2}{*}{$\Treg$} & Left & $-3.0 \pm 0.2$ & $0.2 \pm 0.2$ \tabularnewline
                          & Right & $-3.3 \pm 0.2$ & $0.1 \pm 0.2$ \tabularnewline
\rule{0pt}{3ex}
\multirow{2}{*}{$\Tirr$} & Left & $-3.5 \pm 0.2$ & $-0.1 \pm 0.2$ \tabularnewline
                          & Right & $-3.6 \pm 0.2$ & $0.1 \pm 0.2$ \tabularnewline
\hline 
\end{tabular}
\end{center}
\end{table}

\subsection{The effects of prior choices in searches for Bianchi signatures} \label{section:prior_effect}

\begin{figure}
 \centering
 \includegraphics[width=0.5\textwidth]{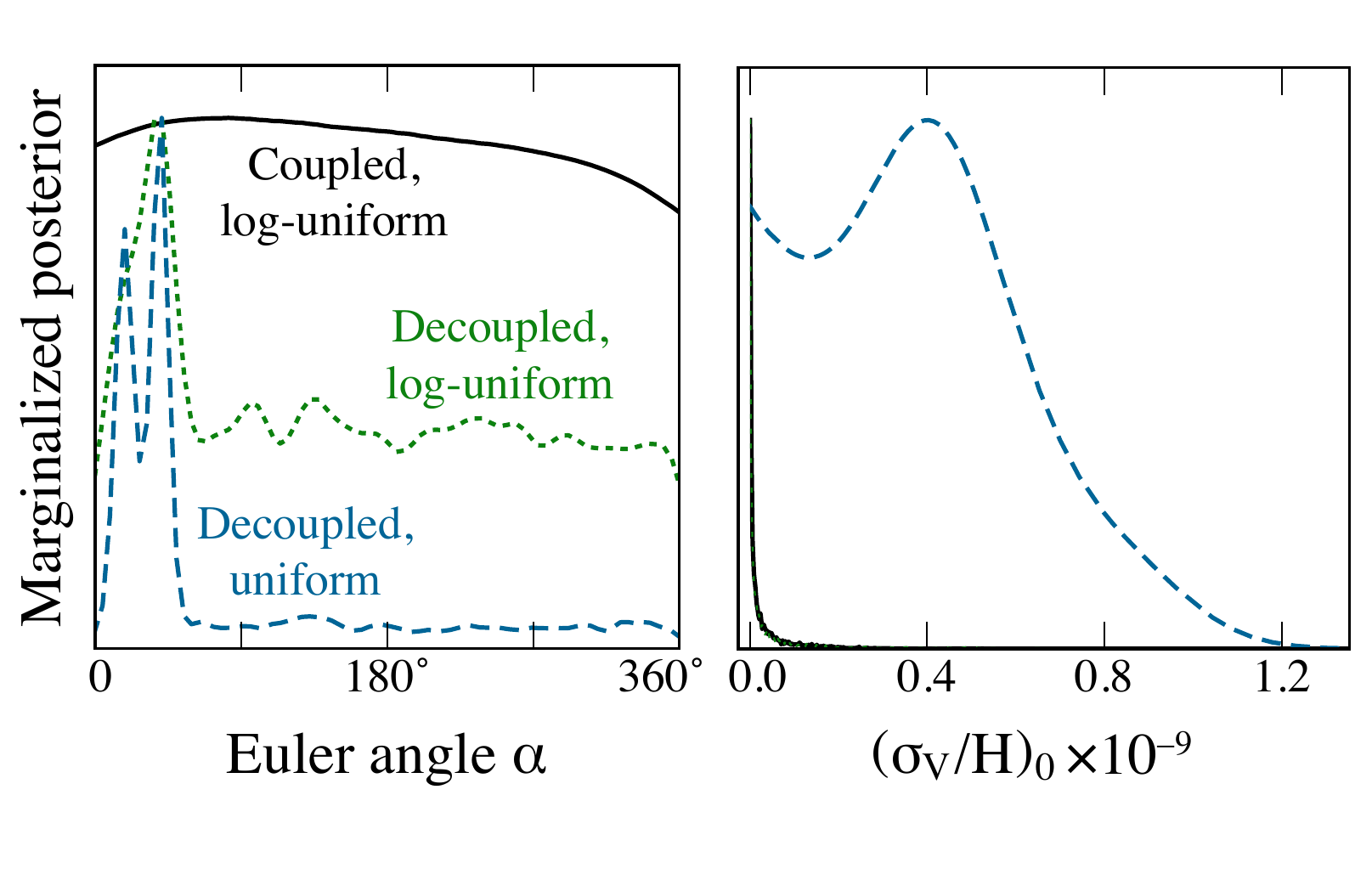}
 \caption{The strong prior-dependence of Bianchi template correlations is illustrated using posterior distributions on Euler angle $\alpha$ and shear $\sigmaVH$ for (black solid lines) coupled model with a log-uniform prior on $\sigmaVH$, (green dotted lines) decoupled model with a uniform prior on $\sigmaVH$, and (blue dashed lines) decoupled model with a uniform prior on $\sigmaVH$.  The last of these hints at the presence of anisotropy, since the position of the anisotropic features (from $\alpha$ and the other Euler angles) begin to be constrained and significant shear appears to be permitted. However this prior is the least motivated physically and statistically. } \label{Fig:prior_effect}
\end{figure}

Figure \ref{Fig:prior_effect} shows the posterior distributions recovered for the Bianchi parameters $\sigmaVH$ and $\alpha$ in searches for $\V$ modes in the WMAP ILC map, assuming the following prior choices:
\begin{itemize}
 \item $\Vc$, log-uniform prior on $\sigmaVH$ (solid black line);
 \item $\Vd$, log-uniform prior on $\sigmaVH$ (dotted green line);
 \item $\Vd$, uniform prior on $\sigmaVH$ (dashed blue line).
\end{itemize}
The parameter $\sigmaVH$ controls the amplitude of the Bianchi component while $\alpha$ partially controls its orientation. The remaining Euler angles $\beta$, $\gamma$ and the spiral parameter $x$ exhibit similar behaviour to $\alpha$.

{In the log-uniform coupled case (solid black line) we recover the previously-quoted upper limit on shear. On the linear scale of Fig.~\ref{Fig:prior_effect}, the posterior is very sharply peaked towards zero shear (right panel) and there is no preference for orientation (left panel). By decoupling the parameters (dotted green lines), we find that a preference for a particular orientation begins to emerge, in agreement with \citet{Jason}. However, we additionally find that non-zero shear at the amplitude corresponding to the \cite{Jaffe_et_al_2005} template is only permitted once we also switch to a uniform prior on $\sigmaVH$ (dashed blue line). The strong prior-dependence of the analysis shows that even in the case that the parameters are allowed to decouple the data do not robustly support the addition of a Bianchi signal.  } 

\section{Conclusions} \label{section:Conclusions}

{We have presented a new framework to search for departures from background isotropy in CMB data. Our approach extends the \texttt{ANICOSMO} tool in two ways. First, we implemented a new Boltzmann hierarchy solution (\texttt{ABSolve}) to calculate CMB temperature and polarization patterns arising from anisotropy in the background; this allows a much wider variety of solutions to be probed and in future will allow for the inclusion of CMB polarization data in a coherent way. The wider variety of modes in the new solutions require us to adopt shear, rather than vorticity, as the primary parameter. Second, we improved the statistical approach; in particular, we include the previously-neglected effects of the anisotropic background on modes at $\ell>32$.}

{As a test of the new approach and to compare with previous results, we searched for departures from isotropy in WMAP temperature data. In doing so, we included three hitherto unconstrained anisotropy modes, including the regular tensor solution which (unlike other modes) is compatible with exiting inflation in a highly isotropic state. Our setup focuses on Bianchi type VII$_h$ and is easily applicable to other types since these are found by allowing the VII$_h$ parameters to limit to boundary values; in this work, we specifically included VII$_0$ models which are obtained as $\Omega_K \to 0$. }

{We find no evidence for anisotropic expansion from WMAP data and place upper limits on present-day shear, as reported in Table \ref{Tab:results}. Our constraints on vector modes (linked to vorticity) are tighter than those presented in prior work by a factor of five, which we showed to be due to a combination of different priors and our improved treatment of small-scale power. Scalar modes are constrained for the first time at a similar level. However, the first constraint on tensor shear -- and in particular the regular solution to the tensor anisotropy -- is much weaker than the constraint we are able to obtain on the other modes. We showed that this difference arises from the different dynamical nature of the solutions.}
 
{In the near future we plan to extend this framework to allow for analysis of CMB polarization data in addition to temperature. This is expected to further tighten limits on the anisotropy of the Universe~\citep{P&C07}.}

\section*{Acknowledgements}
DS thanks Sabino Matarrese, Alessandro Renzi and Boris Leistedt for helpful discussions.
DS is supported by the Perren Fund and the IMPACT fund and was partially supported by the RAS Small Grant Scheme. SMF is supported by the Science and Technology Facilities Council in the UK. AP is supported by the Royal Society. HVP was partially supported by the European Research Council under the European Community's Seventh Framework Programme (FP7/2007-2013) / ERC grant agreement number 306478-CosmicDawn. JDM was partially supported by the Engineering and Physical Sciences Research Council (grant number EP/M011852/1).
We acknowledge use of the following publicly available codes: \texttt{RECFAST} \citep{RECFAST,RECFAST2};
\texttt{HEALPix}/\texttt{healpy} {\citep{Healpix}}; \texttt{CAMB} \citep{CAMB}; \texttt{MultiNest} \citep{MultiNest1, MultiNest2}; \texttt{S2}\footnote{\url{http://www.jasonmcewen.org/codes.html}} \citep{S2}. We acknowledge use of the Legacy Archive for Microwave Background Data Analysis (LAMBDA). Support for LAMBDA is provided by the NASA Office of Space Science.

\bibliography{Bianchi}

\appendix

\section{Approximations in the high-$\ell$ likelihood} \label{section:Appendix_Cl_approximation}

In this Appendix, we discuss how we have incorporated a correction for the background Bianchi power in high-$\ell$ likelihoods.  For motivation and an overview, see Sec.~\ref{section:statistical_analysis}. In brief, at $\ell>32$ we incorporate the Bianchi power by calculating the sky-averaged equivalent power spectrum $C_{\ell}^{\mathrm{B}}$ and adding it to the isotropic stochastic power  $C_{\ell}^{\mathrm{\Lambda CDM}}$ before passing to a standard high-$\ell$ likelihood.

To understand why this produces a reasonable approximation to the true likelihood, let the measured CMB fractional temperature be expanded in complex spherical harmonics as
\begin{equation}
 \frac{\Delta T}{T}(\vartheta,\varphi)=\sum_{\ell=0}^{\ell_{\text{max}}} d_{\ell m} Y_{\ell m}(\vartheta,\varphi) \, ,
\end{equation}
and similarly for the Bianchi template
\begin{equation}
 \left(\frac{\Delta T}{T}\right)^{(\text{B})}(\vartheta,\varphi)=\sum_{\ell=0}^{\ell_{\text{max}}} b_{\ell m} Y_{\ell m}(\vartheta,\varphi) \, .
\end{equation}

We furthermore define the data and Bianchi power spectra $\hat{C}_{\ell}$ and $C_{\ell}^{\text{B}}$ as 
\begin{align}
 \hat{C}_{\ell} & \equiv\frac{\sum_{m=-\ell}^{\ell}d_{\ell m}^*d_{\ell m}}{2\ell+1} \, , \\
 C_{\ell}^{(\text{B})} & \equiv\frac{\sum_{m=-\ell}^{\ell}b_{\ell m}^*b_{\ell m}}{2\ell+1} \, ,
\end{align}
noting that, unlike in the $\LCDM$ case, the Bianchi power spectrum does not provide lossless data compression (which is to say that the true likelihood cannot be expressed purely in terms of $C_{\ell}^{(\mathrm{B})}$). 

The probability of obtaining the dataset $\mathbf{d}=\{d_{\ell m}, 2\leq\ell\leq\ell_{\text{max}}\}$ given the model, $\M$, is 
\begin{equation}
 P(\mathbf{d}|\mathbf{b},C_{\ell}^{(\Lambda\text{CDM})},\M)\equiv\prod_{\ell=2}^{\ell_{\text{max}}}\mathcal{L}_{\text{exact}}^{(\ell)} \, ,
\end{equation}
with
\begin{equation}
 \begin{split}
 \mathcal{L}_{\text{exact}}^{(\ell)}\equiv&\frac{2^{\ell}}{\left(2\pi C^{(\Lambda\text{CDM})}_{\ell}\right)^{\frac{2\ell+1}{2}}}\times\\
 &\times\exp{\left\lbrace-\frac{(d_{\ell 0}-b_{\ell 0})^2+ {2}\sum_{m=1}^{\ell}|d_{\ell m}-b_{\ell m}|^2}{2C_{\ell}^{(\Lambda\text{CDM})}}\right\rbrace},
 \end{split}
\end{equation}
which is the explicit form of Eq.~\eqref{Eq:lowl_likelihood} for complex spherical-harmonic coefficients.

A typical high-$\ell$ likelihood code assumes a pure $\LCDM$ sky, which in the limit of a full-sky approximation may be written
\begin{equation}
 P(\mathbf{d}|C_{\ell}^{(\Lambda\text{CDM})},\Lambda\text{CDM})\equiv\prod_{\ell=2}^{\ell_{\text{max}}}\mathcal{L}_{\Lambda\text{CDM}}^{(\ell)} \, ,
\end{equation}
with
\begin{equation}
\mathcal{L}_{\Lambda\text{CDM}}^{(\ell)}\equiv\frac{2^{\ell}}{(2\pi C_{\ell})^{\frac{2\ell+1}{2}}}\exp{\left\lbrace-\frac{2\ell+1}{2}\frac{\hat{C}_{\ell}}{C_{\ell}}\right\rbrace} \label{Eq:likelihood_approx}
\end{equation}
and where $C_{\ell}$ is the input theoretical power spectrum.
Our treatment of high multipoles amounts to setting $C_{\ell}=C_{\ell}^{(\text{B})}+C_{\ell}^{(\Lambda\text{CDM})}$. We regard the resulting likelihood as an approximation to Eq.~\eqref{Eq:lowl_likelihood}; it may be written
\begin{equation}
 P(\mathbf{d}|C_{\ell}^{\text{(B)}},C_{\ell}^{(\Lambda\text{CDM})},\M)\equiv\prod_{\ell=2}^{\ell_{\text{max}}}\mathcal{L}_{\text{approx}}^{(\ell)}  \, ,
\end{equation}
with
\begin{equation}
\begin{split}
 \mathcal{L}_{\text{approx}}^{(\ell)}\equiv\frac{2^{\ell}}{\left[2\pi\left(C_{\ell}^{(\text{B})}+C_{\ell}^{(\Lambda\text{CDM})}\right)\right]^{\frac{2\ell+1}{2}}}&\times\\
 &\exp{\left\lbrace-\frac{2\ell+1}{2}\frac{\hat{C}_{\ell}}{C_{\ell}^{(\text{B})}+C_{\ell}^{(\Lambda\text{CDM})}}\right\rbrace} \, .
 \end{split}
\end{equation}
To assess the validity of the approximation, we consider the variance of the quantity
\begin{equation}
 \epsilon^{(\ell)}\equiv\frac{\mathcal{L}^{(\ell)}_{\text{exact}}-\mathcal{L}^{(\ell)}_{\text{approx}}}{\mathcal{L}^{(\ell)}_{\text{exact}}} \, ,
\end{equation}
which has mean zero by the normalization of the likelihoods. If $C_{\ell}^{(\text{B})}>C_{\ell}^{\Lambda\text{CDM}}$, then $\operatorname{var}(\epsilon^{(\ell)})$ can be shown to be infinite (i.e., the approximation can be arbitrarily wrong). However, this is generally not the case, since the Bianchi $C_{\ell}$s tend to be much smaller than the $\LCDM$ $C_{\ell}$s. For $C_{\ell}^{\text{B}}<C_{\ell}^{\Lambda\text{CDM}}$, we have
\begin{equation}
\begin{split}
  &\operatorname{var}\left(\epsilon^{(\ell)}\right)=\left\langle\left(\frac{\mathcal{L}_{\text{approx}}^{(\ell)}}{\mathcal{L}_{\text{exact}}^{(\ell)}}\right)^2\right\rangle-1=\\
  &=\left[\frac{\left(C_{\ell}^{\Lambda\text{CDM}}\right)^2}{\left(C_{\ell}^{\Lambda\text{CDM}}\right)^2-\left(C_{\ell}^{\text{B}}\right)^2}\right]^{\frac{2\ell+1}{2}}\times\exp{\left\lbrace\frac{(2\ell+1)C_{\ell}^{\text{B}}}{C_{\ell}^{\Lambda\text{CDM}}-C_{\ell}^{\text{B}}}\right\rbrace}-1\, ,
\end{split}
\end{equation}
where the angular brackets indicate averaging over all the realizations.

In the limit $C_{\ell}^{(\text{B})} \ll C_{\ell}^{\Lambda\text{CDM}}$, we obtain
\begin{equation}
 \operatorname{var}(\epsilon^{(\ell)})\approx(2\ell+1)\frac{C_{\ell}^{\text{B}}}{C_{\ell}^{\Lambda\text{CDM}}} \, ,
\end{equation}
which is first order in $C_{\ell}^{(\text{B})} / C_{\ell}^{\Lambda\text{CDM}}$. Models such that $C_{\ell}^{(\text{B})} > C_{\ell}^{\Lambda\text{CDM}}$ for some $\ell$ are excluded during the sampling as they fall outside the range within which this approximation is valid -- they constitute, however, a small fraction of the number of explored models.

\bsp

\end{document}